%% file: Si_diffusion.tex
\documentclass[superscriptaddress,aps,prb,latterpaper,showpacs,floatfix]{revtex4}
\usepackage{graphicx}

\def\be{\begin{equation*}}   \def\ee{\end{equation*}}
\def\bea{\begin{eqnarray}}  \def\eea{\end{eqnarray}}

\include{macros}
\begin{document}
%
\title{Self-diffusion on Si(001) mono-hydride surfaces revisited: The role of adatom clustering}
\author{Gefei Qian}
\affiliation{Department of Physics, University of Illinois at Urbana-Champaign, Urbana, IL  61801}
\affiliation{Department of Physics, Illinois State University, Normal, IL 61790}
\author{Xuan Luo}
\affiliation{National Renewable Energy Laboratory, Golden, CO 80401}
\author{S. B. Zhang}
\affiliation{National Renewable Energy Laboratory, Golden, CO 80401}
\author{Yia-Chung Chang}
\affiliation{Department of Physics, University of Illinois at Urbana-Champaign, Urbana, IL  61801}
\author{Dehuan Huang}
\affiliation{Edward L. Ginzton Laboratory, Stanford University, Stanford, CA 94305}
\author{Shang-Fen Ren}
\affiliation{Department of Physics, Illinois State University, Normal, IL 61790}
\affiliation{Edward L. Ginzton Laboratory, Stanford University, Stanford, CA 94305}
%
\begin{abstract}
First-principles total-energy calculations of the H/Si(001)-2x1
surfaces reveals a dual diffusion process for the Si adatoms:
single along the dimer row while pairing up across the row. The
calculated diffusion barrier along the dimer row is 1.1 eV, which
is, however, too small to account for the hydrogen-induced growth
disruption seen by experiments. Instead, we find that the adatom
diffusion, in the presence of H, leads to the formation of
immobile fourfold-ring Si tetramers which are difficult to break.
This could explain the adverse effects of H on Si homoepitaxy.

\end{abstract}
\maketitle
%

\hspace*{4ex}
Silicon diffusion on Si(001) surface has attracted tremendous
interest\cite{brocks,yama,swart,lee1,boro} due to its importance
in silicon single crystal growth. Hydrogen adsorbed Si(001) surface is
the typical starting template for both high temperature and low temperature
growth of silicon devices, latter of which is of particular importance
for nanostructure growth\cite{Tucker}. A recent experiment\cite{Tromp}
shows that with one monolayer (ML) coverage of hydrogen the silicon
growth disrupts at low temperature ($< 200^\circ$C), which can be
resumed at a much higher temperature. This new phenomenon stimulates a
great deal of research interest on the Si mono-hydride
surface\cite{vasek,ogit,Ohno,Oshiyama,hong,smil,lee}. Currently, there are
three possible explanations to this phenomenon. The first one\cite{Tromp},
which is based on the experimental evidence, attributes the interruption
of epitaxy growth to the formation of Si dihydrides on the surface. The
second one\cite{ogit}, which is based on first-principles calculations,
suggests that hydrogen atoms hardly segregate from the surface, hence
disrupting the growth. The third one\cite{Ohno,Oshiyama,lee}, also based
on first-principles calculations, suggests that hydrogen can spontaneously
segregate from the surface. However, regarding the growth disruption, the
calculations differ considerably: while Refs.\cite{Oshiyama,lee} suggest
low barriers implying no disruption, Ref.\cite{Ohno} predicts disruption
due to large barriers. Growth disruption is also related to the surface
roughness caused by the hydrogen adsorption\cite{Adams}, but the explanation
to the roughness varies. In the first explanation the roughness is caused by
higher-order hydrides, as supported by a first-principles calculation\cite{rebo},
which can be overcome at about 200$^\circ$C.

In this paper, we perform a first-principles total-energy
calculation on this problem. We predict that Si adatoms diffuse in
two qualitatively different ways: as single atoms parallel to the
surface dimer row with a 1.1-eV energy barrier but as pairs
perpendicular to the dimer row. The formation of the four-Si
tetramers is predicted to occur during epitaxial growth. Its
effects on the homoepitaxial and heteroepitaxial growth of Ge on
Si are elucidated.

Our calculations are based on the density functional
theory(DFT)\cite{dft} within the local density approximation(LDA)\cite{lda}.
We use a norm-conserving pseudo-potential\cite{Norm} and the
plane-wave basis set to perform the molecular dynamics simulation
via the Car-Perrinello scheme\cite{CP}. The surface is modeled by
a p(4x4) surface cell consisting of six Si layers and a nine-layer
thick vacuum. The lower surface is passivated by hydrogen. While
the bottom two Si + H layers are fixed, the rest of the atoms are
fully relaxed. We used an energy cutoff of 10 Ry with one special
k-point for the supercell Brillouin zone integration. Increasing
the cutoff to 12 Ry or adding sampling k points only change the
total energy by less than 0.1 eV/atom, while the change in the
energy difference between configurations is less than 0.01
eV/atom.

\section{Single adatom adsorption and diffusion}
\begin{figure}[!p]
\begin{center}
\includegraphics[width=5.0in]{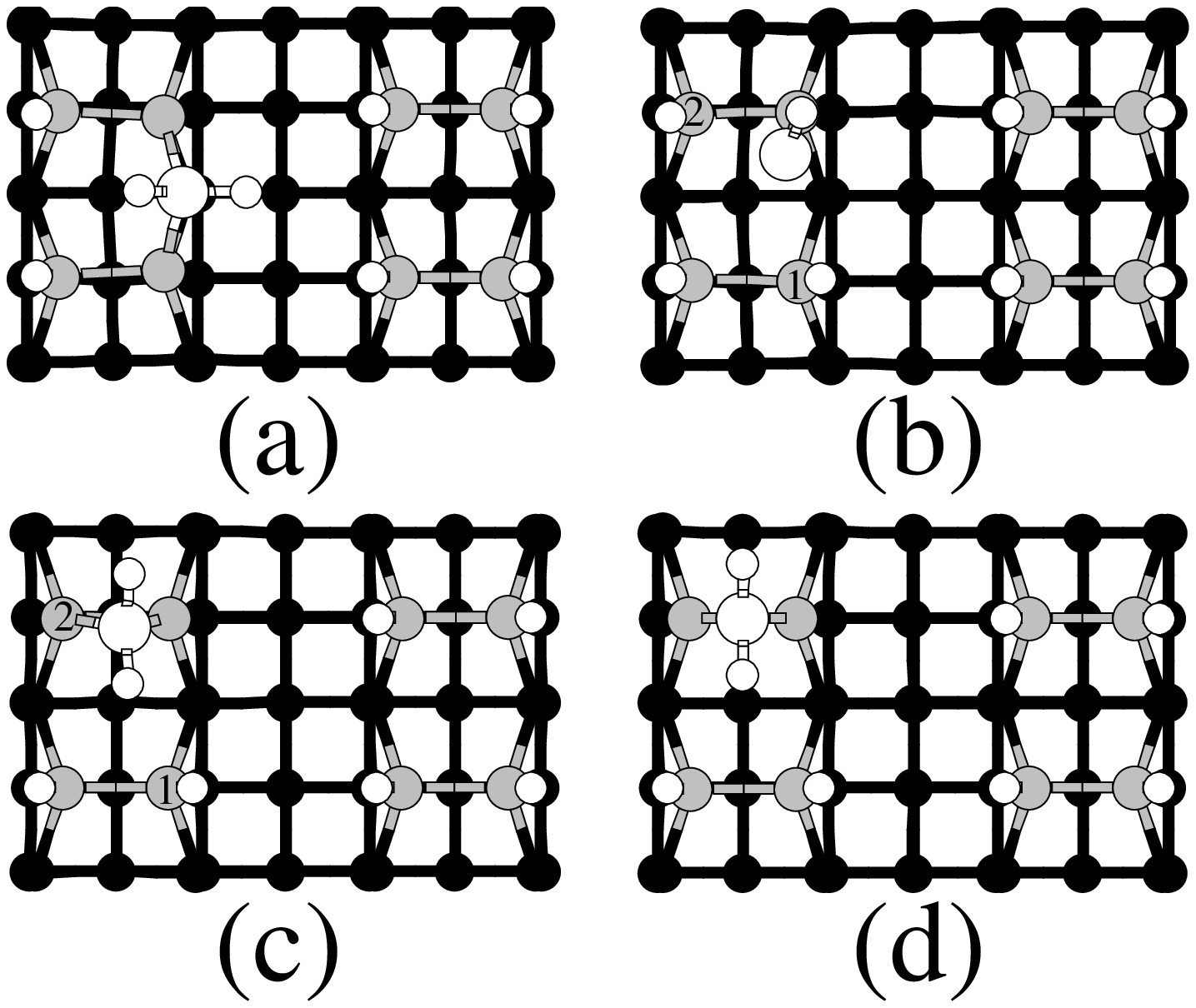}
\caption[Surface structures for Si adatom diffusion]
{ Various surface structures for Si adatom diffusion.
} \label{fig:si}
\end{center}
\end{figure}

.
In our calculation, we mimic actual growth process by placing a Si
adatom above the H/Si(100) surface and then let it move. We find
that the adatom slowly goes to one of the two lowest energy sites
shown in \fig{si}(a) and (d). In qualitative agreement with the
previous calculations by Nara et al.\cite{Ohno}, Jeong and
Oshiyama\cite{Oshiyama}, and Lee et al.\cite{lee}, the capture of
H by the Si adatom from the hydrogenated surface is spontaneous.
However, different from their studies, we find that in both cases,
the Si adatom causes spontaneously segregation of, {\it not one},
but {\it two} hydrogen atoms from the nearby Si-Si dimers to form
highly stable Si-H$_2$ complexes with the adatom. This finding is
consistent with a recent experiment\cite{Shen}, in which it was
shown that most of the H atoms remain to be on top of the surface
during homoepitaxial growth.

While the atomic structures in \fig{si}(a) and (d) resemble that of
Si adatom on the bare\cite{brocks} and on hydrogenated
surfaces\cite{hong}, they differ considerably from
Refs.\cite{Ohno,Oshiyama,lee}, in which the Si adatom diffusion
pathways have been calculated. In the work of Jeong and
Oshiyama\cite{Oshiyama} and Lee et al.\cite{lee}, the Si adatoms
assume only high-energy configurations (as determined from the
present calculations) that involve either one or zero H atom. Not
surprisingly, the calculated diffusion barriers are low (0.7 and
1.0 eV, respectively, in Ref.\cite{Oshiyama} for diffusions
parallel and perpendicular to the surface dimer row). In contrast,
Nara et al.\cite{Ohno} considered a different diffusion pathway
involving a two-step process, intra and inter (1x1) cell hopping
where the capture and release process of the
hydrogen atoms also plays a key role. It was concluded that while
an H$_2$ capture significantly lowers the energy of the Si
adatoms, it is advantageous to move the adatom around within the
(1x1) cell with only one H atom. The calculated intra-cell barrier
is 0.5 eV. However, only one of the low-energy configurations in
\fig{si}, i.e., (d), was considered in the calculation. As a result,
diffusion out of the (1x1) cell (inter-cell hopping) is difficult.
Nara et al. suggested\cite{Ohno} that to diffuse across cells, the
Si adatom must temporarily loss both of the H atoms to become bare
silicon. This unfavorable configuration leads to, on top of the
0.5-eV intra-cell barrier, the unphysical large inter-cell barriers
of 1.0 and 1.2 eV, respectively, for diffusions parallel and
perpendicular to the dimer row.

In the present study, we consider adatom diffusion along a path
connecting the low-energy H$_2$-capture configurations shown in
\fig{si}(a) to (d). The edge-Si adatom in \fig{si}(a) is the ground
state configuration. Interestingly, because this structure has a
mirror symmetry in the ($\bar 1$10) plane, inter- and intra-cell
diffusions make no difference, in contrast to the findings of Nara
et al.\cite{Ohno}. \fig{si}(b) shows a transition state where the
displaced Si adatom gives back one of the H atoms to the surface
atom, labeled 1. The energy is 0.5-eV higher than the ground
state due to the ``quasi" twofold coordination. \fig{si}(c) shows
another transition state where the Si adatom recaptures a second H
from the surface atom, labeled 2. The diffusing adatom eventually
lands on top of the Si-Si dimer in \fig{si}(d), which is only 0.03 eV
higher in energy than the ground state. The Si adatom can diffuse
further up along the dimer row in a reversed order of (d) to (a),
and so on. The calculated upper bound for the diffusion barrier is
0.7 eV.

With the 0.7-eV barrier, a growth temperature of 400K, and with the typical vibration frequency
for the surface atoms of $10^{13}H_z$\cite{boro}, we estimate that the diffusion event happens at
a frequency of $2\times10^4H_z$.
The Si adatom is ready to diffuse  along the
dimer row. On the other hand, diffusion perpendicular to the dimer row is much more difficult.
There are two reasons for this: (i) diffusion across the dimer rows requires the adatom to give
back both H atoms, which is energetically costly, 
and (ii) in the middle of the trough, the adatom
is to a large extent unbonded.

\section{Formation of the adatom clusters}.
The above discussion shows that Si adatom diffusion rate is not
the reason for the disruption of silicon homoepitaxy growth with
1ML H coverage\cite{Tromp}. Instead, Si adatoms should readily
diffuse, at least parallel to the dimer row. This leads to the
possibility of adatom clustering, which will be considered next.

\begin{figure}[htbp]
\begin{center}
\includegraphics[width=5.0in]{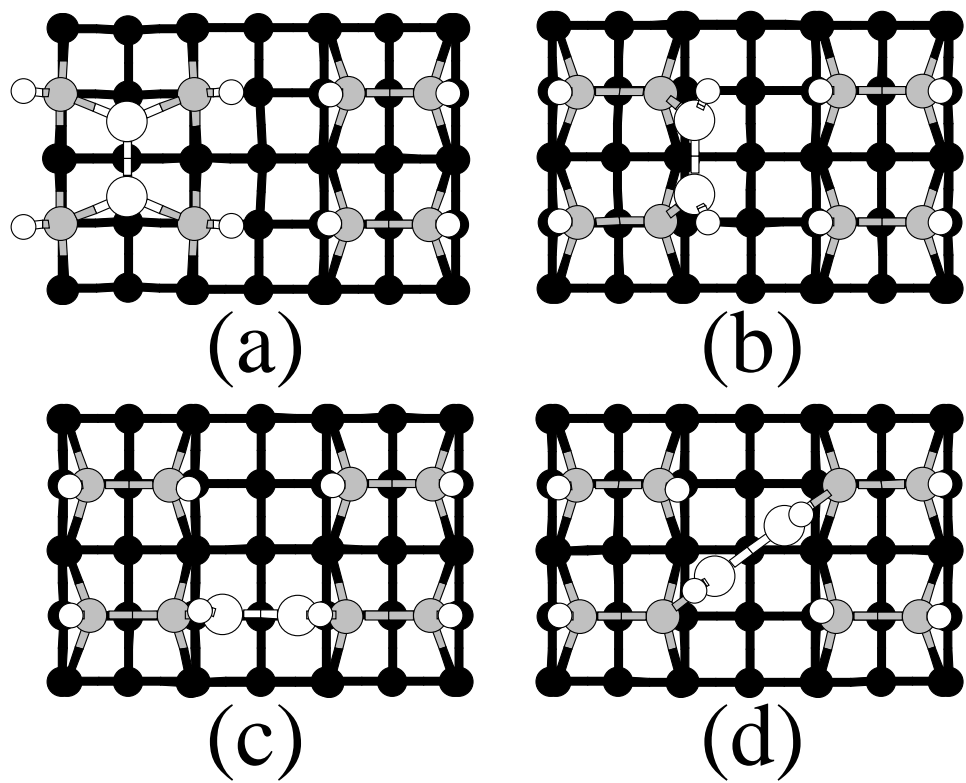}
\caption[Si$_2$ diffusion]
{ Various surface structures for Si$_2$ diffusion with
total energy (in eV/adatom): (a)E=-0.19,
(b)E=-0.24,(c)E=-0.19,(d)E=-0.14.} \label{fig:si2}
\end{center}
\end{figure}

\fig{si2} shows the calculated low-energy Si-adatom pairs along
with their energies. Throughout the paper, the energy of the
edge-Si adatom in \fig{si}(a) is taken as the energy zero. Given the
negative energies in \fig{si2}, it is clear that the formation of the
pairs is energetically favored. The atop-dimer in \fig{si2}(a) is
typical of the surface dimers, whereas the edge-dimer and
in-trough dimer in \fig{si2} (b) and (c) are not, although they have
lower or comparable total energy. A closer examination of the
latter reveals that in both cases the Si ad-dimer is engaged in a
planar $sp^2$ configuration with nearly 120$^\circ$ bond angles,
not seen on bare silicon\cite{lee1,boro,zhang,bedr}.

Recently, it has been suggested that adatoms may diffuse in pairs
with a rate comparable to or even faster than individual
adatoms\cite{lee1,boro}. Here, we study such a possibility.
Consider the edge-dimer configuration in \fig{si2}(b): In order to
become the in-trough dimer in \fig{si2}(c), it is required to go
through an intermediate step as shown in \fig{si2}(d). Note that the
ad-dimer diffusion parallel to the dimer row is simply a repeat of
the pattern in \fig{si2}, i.e.,
(b)$\rightarrow$(d)$\rightarrow$(c)$\rightarrow$(d)$\rightarrow$(b),
$\cdots$ etc. The corresponding energy barrier is 1.2 eV. It is
also interesting to note that while diffusion perpendicular to the
dimer row is difficult for single adatoms, it could be much easier
for ad-dimers, which could be treated separately as (i) diffusion
across the dimer row and (ii) diffusion across the trough. Process
(i) can be realized via the intermediate step in \fig{si2}(a), which
requires only the rotation of the ad-dimer with respect to the
surface atoms and a transfer of the H atoms. Process (ii) is no
different from diffusion parallel to the dimer row, with the
resulting edge-dimer, however, on the other side of the trough.
The corresponding diffusion barrier is 1.2 eV.

\begin{figure}
\begin{center}
\includegraphics[width=5.0in]{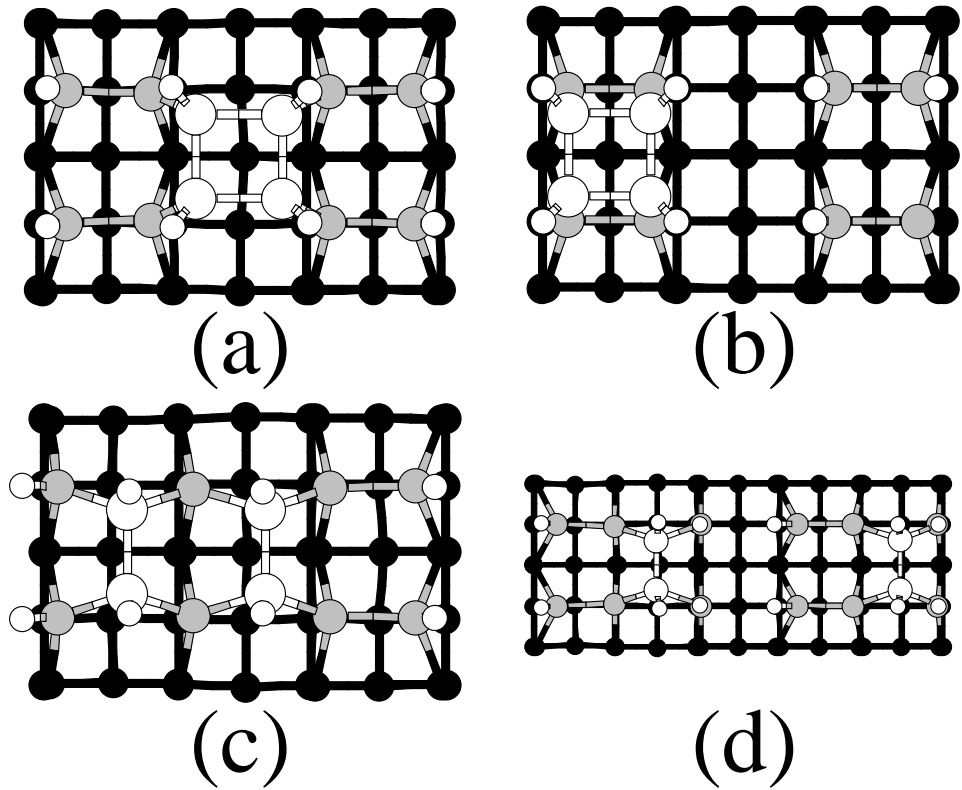}
\caption[Si$_4$ clustering]
{ Various surface structures for Si$_4$ clustering with
total energy (in eV/adatom): (a)E=-1.02,
(b)E=-0.80,(c)E=-1.09,(d)E=-0.83. } \label{fig:si4}
\end{center}
\end{figure}

Speaking of clusters, it is also important to consider the
4-adatom clusters or the tetramers. \fig{si4} shows four such
configurations made of the ad-dimer pairs. These clusters can be
classified into three different categories: (i) ad-dimer pairs with
direct bonding such as those in \fig{si4}(a) and (b), (ii) ad-dimer
pairs in close association but without any direct bonding such as
the one in \fig{si4}(c), and (iii) ad-dimer pairs distant apart. All
the tetramers here have energies considerably lower than either
the adatom or ad-dimer, for example, by as much as 1.09 eV per
adatom.

In case (i), the adatoms form fourfold-ring square structures with
90$^\circ$ bond angles. With respect to the surface atoms, the
trough-tetramer in \fig{si4}(a) forms five- [along the ($\bar 1$10)
direction] and seven-fold [along (110)] rings, respectively. On
the other hand, the atop-tetramer in \fig{si4}(b) forms five- [along
($\bar 1$10)] and four-fold [along (110)] rings, respectively.
Naturally, the latter (-0.8 eV/atom) have significantly higher
energy than the former (-1.02 eV/atom). In case (ii), the two
ad-dimers form a second nearest neighbor (NN) pair as shown in 
\fig{si4}(c). These are dimers similar to those on the substrate, except
being one layer higher and rotated by 90$^\circ$. The NN-tetramers
have the lowest energy because they involve only the five- [along
($\bar 1$10)] and six-fold [along (110)] rings. The energy is not
much lower than the trough-tetramers (only by 0.07 eV per Si),
however, because of the strain, which can be easily seen by the
large adatom displacements with respect to the surface atoms in
\fig{si4}(c). In case (iii), two distant ad-dimers are bound with an
energy, -0.83 eV per Si through flipping surface dimers in
between. \fig{si4}(d) shows the distant tetramers with the smallest
ad-dimer separation possible.

\begin{figure}
\begin{center}
\includegraphics[width=5.0in]{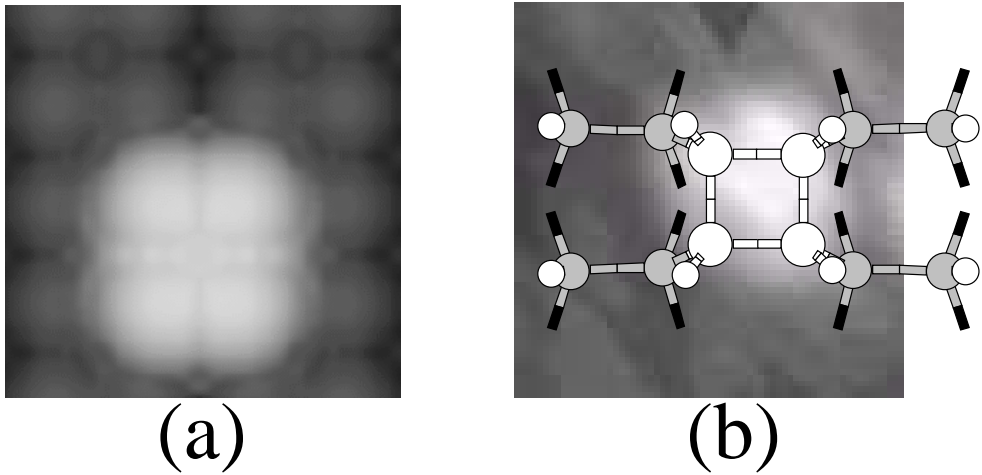}
\caption[Simulated and Experimental STM image]
{ Comparison of (a) the simulated STM image to (b)the
experimental STM image  } for the trough tetramer.\label{fig:stm}
\end{center}
\end{figure}

According to their effects on the homoepitaxial growth, we can
also classify the tetramers in terms of being a growth mode or
growth-interruption mode. By this definition, the NN- and the
distant tetramers belong to the growth mode because they are
consistent with the further growth of the adatom clusters into a
new surface layer indistinguishable from the substrate. The
trough-tetramers, on the other hand, belong to the
growth-interruption mode, as here two ad-dimers are packed into a
region with twice of the normal adatom density. To resume the
growth from the growth-interruption mode, either one has to break
up the trough-tetramers, which will cost energy, or the growth
front will become defective. Recently, it has been
shown\cite{spitz} that the smallest stable adatom islands in the
CVD growth (in which H is involved) consist of two dimers or a
tetramer. \fig{stm} compares the calculated STM image for the
trough-tetramer with experiment where adatom clusters are created
by applying electrical pulse through STM tip coated with silicon.
The relative stability of the clusters, the size and the relative
position of the images with respect to the substrate suggests that
the trough-tetramer model could explain the experiment.

Finally, although the NN-tetramer has a slightly lower energy than the trough-tetramer (by 0.07 eV/atom),
kinetically, the trough-tetramers are more readily to form because the diffusing adatoms and ad-dimers are
mostly in the forms of edge-atoms and edge-dimers, respectively.
Hence, the trough-tetramers have a much
higher probability to form than the NN-tetramers whose formation is an activated process in order to move
the adatoms/ad-dimers from the trough region to atop the dimer row.
This kinetic enhancement, together with the
difficulty to break up the fourfold rings, provides an alternative explanation
to the growth disruption by H.

Because part of the surface dimers in the growth mode, either the
NN-tetramer in \fig{si4}(c) or the distant tetramer in \fig{si4}(d), is
strained, replacing the Si adatoms by Ge adatoms, or replacing
some of the surface Si atoms by Ge as suggested by a recent
study\cite{rudk}, could lower the strain energy because Ge is
larger and prefers longer bonds. Hence, one can expect that the
NN- and distant tetramers to be stabilized with respect to the
trough-tetramers during heteroepitaxy on Si substrates. In other
words, while hydrogenation leads to the interruption of the
homoepitaxy, a similar effect could be absent in the heteroepitaxy
of germanium. Indeed, it is known\cite{kahng} that the presence of
hydrogen assists the epitaxial growth of Ge on Si, against the
formation of islands.

\section{Summary}
In summary, we have studied the Si adatom diffusion processes on
the H/Si(100)(2x1) surface by first-principles total-energy
calculations. Our results show that Si adatom diffusion along the
dimer row proceeds with an energy barrier significantly lower than
what is required for growth disruption, whereas diffusion
perpendicular to the dimer row involves the motion of adatom
pairs. Study of the adatom tetramers reveals that the formation of
the growth-interruption trough-tetramers could be the reason for
the hydrogen induced disruption of homoepitaxial growth of
silicon.

We are grateful to T. C. Shen for fruitful discussions. The work
at UIUC was supported in part by DARPA DAAD19-01-1-0324. S. F. Ren
was supported by NSF 9803005 and NSF 0001313. The work at NREL was
supported by the U.S. DOE-SC-BES under contract No.
DE-AC36-99GO10337 and by the NERSC for MPP time.

\end{document}

%% file: macros.tex
\def\ba{\begin{array}}
\def\bea{\begin{eqnarray}}
\def\be{\begin{equation}}
\def\bga{\begin{align}}

\def\dgk12{\Delta G_k^2}
\def\dgk{{\bf G}_\|-{\bf G}_\|\p+{\bf G}_s-{\bf G}_s\p}

\def\ea{\end{array}}

\def\eea{\end{eqnarray}}
\def\ee{\end{equation}}

\def\ena{\end{align}}

\def\flz2{f_{\alpha_2,\nu_2}^{\sigma_2,l_z^2}(z)}
\def\flz{f_{\alpha\nu}^{\sigma l_z}(z)}
\def\flzp1{f_{\alpha_1,\nu_1}^{\sigma_1,l_z^1}(z)}
\def\flzp{f_{\alpha'\nu'}^{\sigma' l'_z}(z)}

\def\p{^\prime}

\def\wan0{W_{n,l_z,\gs+{\bf k}_\|}}

\newcommand{\comment}[1]{}

\newcommand{\fig}[1]{FIG.~\ref{fig:#1}}

\newcommand{\gs}[1]{{\bf G}_{s{#1}}}

\def\bgm{\begin{minipage}{10em}}
\def\edm{\end{minipage}}